\documentclass[aps,prb,11pt,reprint,twocolumn,amsmath]{revtex4-1}
\usepackage[utf8]{inputenc}
\usepackage{graphicx}

\makeatletter
\renewcommand{\p@subsection}{}
\renewcommand{\p@subsubsection}{}
\makeatother

\newcommand{\beq}[0]{\begin{equation}}
\newcommand{\eeq}[0]{\end{equation}}

\begin{document}

\title{Inventory effects on the price dynamics of VSTOXX futures quantified via machine learning}
\author{Daniel Guterding}
\email{daniel.guterding@eurexchange.com}
\affiliation{Eurex, Deutsche Börse AG, 60485 Frankfurt am Main, Germany}

\date{\today}

\begin{abstract}
The VSTOXX index tracks the expected 30-day volatility of the EURO STOXX 50 equity index. Futures on the VSTOXX index can, therefore, be used to hedge against economic uncertainty. We investigate the effect of trader inventory on the price of VSTOXX futures through a combination of stochastic processes and machine learning methods. We formulate a simple and efficient pricing methodology for VSTOXX futures, which assumes a Heston-type stochastic process for the underlying EURO STOXX 50 market. Under these dynamics, approximate analytical formulas for the implied volatility smile and the VSTOXX index have recently been derived. We use the EURO STOXX 50 option implied volatilities and the VSTOXX index value to estimate the parameters of this Heston model. Following the calibration, we calculate theoretical VSTOXX future prices and compare them to the actual market prices. While theoretical and market prices are usually in line, we also observe time periods, during which the market price does not agree with our Heston model. We collect a variety of market features that could potentially explain the price deviations and calibrate two machine learning models to the price difference: a regularized linear model and a random forest. We find that both models indicate a strong influence of accumulated trader positions on the VSTOXX futures price.
\end{abstract}

\maketitle

\section{Introduction}
Volatility derivatives are nowadays in widespread use, since they can provide protection against economic uncertainty.~\cite{carr2009} Furthermore, traders can employ these derivatives to express their sentiment with respect to the expected volatility of financial markets. The volatility benchmark for the European equity market is the VSTOXX index, which reflects the expected volatility over a time horizon of 30 days and is calculated from EURO STOXX 50 option volatilities. The future on this index is the VSTOXX future, which is a very liquid and actively traded product used for hedging purposes or to obtain pure exposure to volatility.

The VSTOXX index is calculated based on EURO STOXX 50 option prices. Therefore, the price dynamics of the VSTOXX future are different from those of equity or equity index futures. The methodology for pricing VSTOXX futures and similar volatility derivatives is still an open question that is actively discussed in the literature.~\cite{carr2009, gruenbichler1996, zhang2006, lin2007, zhu2007, sepp2008, lin2009, lu2009, psychoyios2010, duan2010, lin2010, zhang2010, wang2011, zhu2011, zhulian2012, luo2012, hancock2012, lian2013, cont2013, baldeaux2014, frijns2016, li2017, pacati2018, lo2019} While most of these studies focus on finding the right stochastic process to price volatility derivatives, the combination of these models with modern machine learning techniques has to our knowledge not yet been studied.
Here we investigate to what extent the prices of the VSTOXX future can be explained directly by the EURO STOXX 50 option prices and to which extent other traditionally not considered factors in the trading environment play a role in pricing this contract. 

We first construct a continuous-time stochastic volatility model, namely the Heston model, to obtain a theoretical reference price for the VSTOXX future. We then apply this methodology to the price time series from May 2016 to August 2019. Although the model prices are usually in line with the market, we also observe systematic deviations over extended time spans, which we attempt to explain based on other factors of the trading environment, like aggregated trader positions, using machine learning methods. We conduct an investigation of correlations between these price deviations and pre-selected promising feature variables. Furthermore, we discuss a generalized linear model and a non-linear random forest trained on these features. We analyze the accuracy of our models in describing the observed price deviations. Finally, we assess the explanatory power of the included model features and relate our findings to the economics of trading.

\section{VSTOXX future pricing}
\subsection{Heston stochastic process}
The Heston model is a stochastic volatility model. In such models not only the value of the underlying asset may vary, but also the volatility of the underlying process itself.~\cite{heston1993} These models may account for effects that are missing from the seminal Black-Scholes model~\cite{blackscholes73}, such as skew and smile of the implied volatility. The Heston model is a special instance of a stochastic volatility model, for which semi-analytical pricing formulas for plain-vanilla options exist. It consists of a coupled log-normal spot process and a mean-reverting variance process. The log-spot at time $t$ is denoted as $x_t = \ln (S_t)$, where in our case $S_t$ is the actual EURO STOXX 50 index level. The instantaneous variance is denoted as $v_t$. Using these abbreviations, we write the stochastic differential equation for the Heston model as:
\begin{subequations}
\begin{align}
dx_t &= \Big( \mu - \frac{v_t}{2} \Big) dt + \sqrt{v_t} dW_t^x \\[4pt]
dv_t &= \kappa ( \theta - v_t ) dt + \xi \sqrt{v_t} dW_t^v \\[4pt]
\rho dt &= dW_t^x \cdot dW_t^v 
\end{align}
\label{eq:hestonsde}
\end{subequations}
Here, $W_t^x$ and $W_t^v$ are Wiener processes, $\rho$ is the correlation between those processes, $\mu$ is the drift of the spot process, $\kappa$ is the speed of mean-reversion for the variance process, $\theta$ is the long-term variance and $\xi$ is the volatility of volatility. Furthermore, the initial variance $v_0$ has to be determined, since it is required for the evaluation of pricing formulas in the following sections. 

\subsection{Calibration of the Heston model against plain-vanilla options volatilities}
We determine the parameters of the Heston model by calibrating it against a set of instruments with known prices. Here, we use the daily settlement prices of EURO STOXX 50 options on Eurex. From those we calculate implied volatilities using the method by Jäckel.~\cite{jaeckelrational} For the calibration we use options with a rather long time to expiry. Since EURO STOXX 50 options on Eurex are listed with a fixed expiry date, the remaining time to maturity of the available options naturally decreases as time progresses. We discard all options with a remaining time to maturity of more than 300 days and then select the hindmost remaining expiry date.

We also restrict the strike range of the options by restricting their moneyness to the range $[-14, 5]$, since options that are too far out-of-the-money (OTM) or too far in-the-money (ITM) are less liquid than the at-the-money (ATM) options and tend to decrease the stability of the calibration process when included. The moneyness $m$ can be calculated based on the following expression:
\beq
m = \frac{\ln (F / K)}{\sigma_\text{imp} \sqrt{\tau}}
\eeq
Here, $F$ is the forward price of the underlying, $K$ is the option strike, $\sigma_\text{imp}$ is the implied volatility and $\tau$ is the time to maturity. In place of $\sigma_\text{imp}$, for simplicity we use the ATM implied volatility at the given time to maturity $\tau$ irrespective of the actual strike $K$.

\begin{figure*}[t]
\includegraphics[ width=\linewidth ]{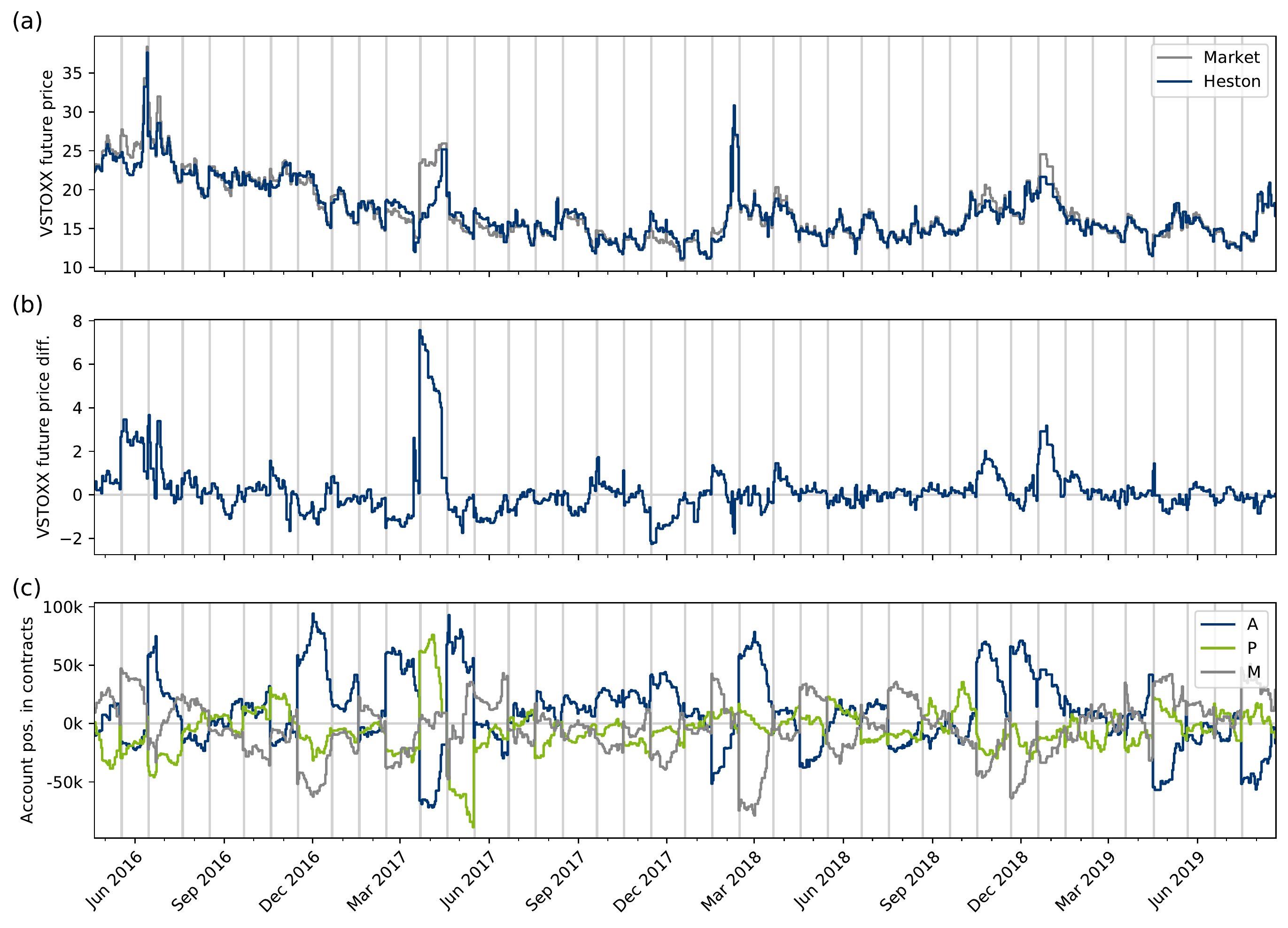}
\caption{(a) Market and Heston model prices for the front month VSTOXX future contract between May 2016 and August 2019. (b) Difference between Market and Heston model price. (c) Aggregated account positions for Agency (A), Proprietary (P) and Market Maker (M) in the front month VSTOXX future contract. Positive values for the position indicate the number of contracts in a long positions, while negative numbers correspond to a short position. The expiry dates of the front month contract are marked by vertical grey lines.}
\label{fig:priceposition}
\end{figure*}

For the implied volatilities $\sigma_H$ of the Heston model as a function of the strike $K$, time to maturity $\tau$ and the model parameters $(\kappa, \theta, \xi, \rho, v_0)$, no exact analytical expression is known. However, for the Heston model the prices of plain-vanilla call and put options can be calculated from a semi-analytical formula involving a numerical integral.~\cite{heston1993} The efficient evaluation of this integral has been the subject of recent research.~\cite{lefloch2018, guterding2018} Here, we use an even more simple approximate analytical expression for $\sigma_H (K, \tau, \kappa, \theta, \xi, \rho, v_0)$, which has been derived from a recently proposed expansion method.~\cite{lorig2017} Since the resulting formulas are rather lengthy, we refer the interested reader to ref.~\onlinecite{lorig2017}. Although the accuracy of this method is somewhat lower compared to the numerically exact semi-analytical formula, it significantly accelerates the calibration process and delivers the Heston parameters with sufficient accuracy.

Based on the expression for the implied volatility smile $\sigma_\text{H} (K, \tau, \kappa, \theta, \xi, \rho, v_0)$ under the Heston model, we can extract the parameters by  fitting the volatility smile observed in the market. During the optimization process the parameters are constrained to the following ranges: $ \kappa \in [0.01, 20]$, $\theta \in [0.01, 1]$, $\xi \in [0.01, 5]$, $\rho \in [-1, 1]$, $v_0 \in [0.01, 1]$. The objective function of the optimization is set to the squared deviation between market implied volatilites $\sigma_{\text{imp},K}$ and model implied volatilities $\sigma_{\text{H}, K} \equiv \sigma_\text{H} (K, \tau, \kappa, \theta, \xi, \rho, v_0)$, weighted by the Black-Scholes vega at the respective strike $K$:
\beq
(\text{MSE})_\sigma = \frac{\sum\limits_K \nu_K  \cdot (\sigma_{\text{imp},K} - \sigma_{\text{H}, K})^2 }{\sum\limits_k \nu_K}
\label{eq:msevola}
\eeq
The Black-Scholes vega measures the sensitivity of the option price with respect to implied volatility, i.e.~market implied volatilities associated with a large vega are more reliable than data points with a small vega. The expression for the vega $\nu$ at strike $K$ is given by:
\begin{subequations}
\begin{align}
\nu_K &= S_0 \sqrt{\tau} \varphi (d_1) \\
d_1 &= \frac{\ln (S_0 / K) + \Big(\mu + \frac{ \sigma_\text{imp}^2}{2} \Big) \tau}{\sigma_\text{imp} \sqrt{\tau}}
\end{align}
\end{subequations}
Here, $\varphi (x)$ is the standard normal density function and $S_0$ is the initial value of the underlying, i.e.~the EURO STOXX 50 index value.

\subsection{Calibration of the Heston model against the VSTOXX index level}
The VSTOXX index quantifies the expected 30-day volatility of the EURO STOXX 50 index. The pricing formula for the VSTOXX index in the Heston model is a special case of a formula for a more general model, which was recently derived in ref.~\onlinecite{zhulian2012}. Within the Heston model, these equations for the VSTOXX index level can be reduced to:
\begin{subequations}
\begin{align}
\bar \tau &= \frac{30}{365} \\
a &= \frac{1 - e^{-\kappa \bar \tau}}{\kappa \bar \tau} \\
b &= \theta (1 - a) \\
(\text{VSTOXX})_t &= 100 \sqrt{ (a v_0 + b)} \label{eq:VSTOXXindex}
\end{align}
\end{subequations}
Therefore, the VSTOXX level is known once $(\kappa, \theta, v_0)$ are known. Similarly, a known value of the VSTOXX index determines the relation between $\kappa$, $\theta$ and $v_0$. We use these equations to calibrate the Heston model so that the VSTOXX index value is reproduced. We define the squared error in the VSTOXX index based on the VSTOXX value $(VSTOXX)_{\text{obs}}$ observed in the market and the VSTOXX value $(VSTOXX)_{\text{H}}$ calculated from the Heston model:
\beq
(SE)_\text{IDX} = \big[ (VSTOXX)_{\text{obs}} - (VSTOXX)_{\text{H}} \big]^2
\label{eq:seindex}
\eeq
Unlike in the case of the volatility smile, the single index value cannot be used to solve for the whole set of Heston parameters. Instead it practically just eliminates one of the parameters $(\kappa, \theta, v_0)$. Therefore, calibration against the VSTOXX index level should be used in conjuction with another calibration step, e.g.~calibration with respect to the smile in implied volatilities.

\subsection{Combined calibration of the Heston model against plain-vanilla options volatilities and VSTOXX index level}
We use the aforementioned error measures introduced in eq.~\ref{eq:msevola} and eq.~\ref{eq:seindex} to minimize the deviation of our Heston model both from the observed volatility smile and the observed VSTOXX index level. We combine these equations into a single weighted error:
\beq
(MSE) = w_\sigma \cdot (\text{MSE})_\sigma + w_\text{IDX} \cdot (SE)_\text{IDX}
\label{eq:mseopt}
\eeq
The values of the weights are chosen empirically, so that both volatility smile and index level are reproduced with acceptable accuracy. We use the weights $w_\sigma = 100^2 = 10000$ and $w_\text{IDX} = 2$. The value of $w_\sigma$ merely compensates for the different scales of volatility and VSTOXX level, i.e.~the factor of 100 introduced in eq.~\ref{eq:VSTOXXindex}. The increased weight of $w_\text{IDX}$ ensures that the VSTOXX index level is actually used to fix the relationship between $(\kappa, \theta, v_0)$ as mentioned in the previous section.

We now apply eq.~\ref{eq:mseopt} to a series of daily data points for the option volatilities and VSTOXX index levels. We seek to minimize the deviation calculated from eq.~\ref{eq:mseopt} to find a parameter set that optimally reproduces the observed market data. For the first day in our time-series we determine the parameters $(\kappa, \theta, \xi, \rho, v_0)$ from a global optimization step via differential evolution~\cite{diffevo} followed by local optimization using the L-BFGS-B method.~\cite{lbfgs} All following calibration runs are performed using the parameter set of the previous trading day as starting point and applying local optimization using the L-BFGS-B method only. 

\subsection{Pricing formula for the VSTOXX future}
Using the determined time series of daily Heston parameters $(\kappa, \theta, \xi, \rho, v_0)$ we price the VSTOXX future following the methodology laid out in ref.~\onlinecite{zhulian2012}. When reduced to the special case of the Heston model, the pricing formula from ref.~\onlinecite{zhulian2012} can be rewritten as:
\begin{subequations}
\begin{align}
\bar \tau &= \frac{30}{365} \\
a &= \frac{1 - e^{-\kappa \bar \tau}}{\kappa \bar \tau} \\[4pt]
b &= \theta (1 - a) \\[4pt]
C (\phi, \tau ) &= -\frac{2 \kappa \theta}{\xi^2} \ln \Big( 1 + \frac{\xi^2 \phi}{2 \kappa} \big(e^{-\kappa \tau} - 1 \big) \Big) \\[4pt]
D (\phi, \tau ) &= \frac{2 \kappa \phi}{\xi^2 \phi + (2 \kappa - \xi^2 \phi) e^{\kappa \tau}} \\[4pt]
f (\phi, \tau) &= \exp \Big( C (\phi, \tau) + D (\phi, \tau) v_0 \Big) \\[4pt]
F (\tau) &= \frac{1}{2 \sqrt{\pi}} \int\limits_0^\infty ds \, s^{-3/2} \Big( 1 - f (-sa, \tau) e^{-2b} \Big) \label{eq:vstoxxfutureintegrand}
\end{align}
\label{eq:vstoxxfutureprice}
\end{subequations}
The final price for the future on the VSTOXX with time to maturity $\tau$ is given by $F(\tau)$ as calculated from eq.~\ref{eq:vstoxxfutureprice}. The futures price can be brought to the same scale as the VSTOXX index by multiplying it with a factor of 100.

At first glance, it seems like the integrand of eq.~\ref{eq:vstoxxfutureintegrand} diverges at the lower boundary, i.e.~for $s \to 0$. After applying l'Hospital's rule, a closer inspection shows that the divergence of the integrand is proportional to $s^{-1/2}$. Therefore, the integral actually converges, although the integrand cannot be evaluated directly at the lower boundary. We evaluate the integral of eq.~\ref{eq:vstoxxfutureintegrand} numerically using the standard trapezoidal rule on a grid containing $10^4$ points evenly spaced on a logarithmic scale from $10^{-12}$ to $10^{20}$. 

\subsection{Comparison between market and theoretical prices for the VSTOXX future}
We calculate the price for the VSTOXX future contract that is closest to expiry, i.e.~the front month contract. We switch to the contract that expires in the following month, i.e.~the back month contract, one day before actual expiry of the front month. The exact date for the switch does not influence the results significantly as long as it stays within a few days to actual contract expiry. 

A comparison of our VSTOXX future prices calculated from the Heston model to the prices observed in the market is shown in fig.~\ref{fig:priceposition}a. The difference between those prices, calculated as the market price minus the theoretical price based on the Heston model ($\Delta F = F_\text{obs} - F_\text{H}$), is shown in fig.~\ref{fig:priceposition}b. 

We observe that the time series of calculated prices follows the market-observed price very closely and that deviations are usually below one point of the VSTOXX future value. This level of pricing accuracy is probably not sufficient for market participants actively trading these products, but allows us to study the large deviations between theoretical and observed price, which occasionally appear for extended periods of time.

These time periods are, among others with minor deviations, June 2016 (where the United Kingdom's referendum to leave the European Union took place), April 2017 (probably related to the French presidential election), November 2017 (possibly related to uncertainty about possible government coalitions after a German federal election), November 2018 and January 2019 (both possibly related to the so-called yelllow vest protests in France).

\begingroup
\squeezetable
\begin{table*}[t]
\begin{ruledtabular}
\begin{tabular}{ll}
Feature name & Description \\
\hline
DiffPrice & The difference between market and model price of the VSTOXX front month future contract. This is the target variable. \\
MarketPrice & The market-observed price of the VSTOXX front month future contract. \\
VSTOXX & The value of the VSTOXX index. \\
FitResidual & The residual of the Heston model fit calculated from eq.~\ref{eq:mseopt}.  \\
DaysToExpiry & The number of days until expiry of the front month future contract. \\
PosA & The aggregated long/short position of A/P/M accounts in the VSTOXX front month future contract. \\
PosP & \\
PosM & \\
PosChangeA & The change in position in the VSTOXX front month future aggregated over all A/P/M accounts compared to the previous day. \\
PosChangeP &  \\
PosChangeM &  \\
TradVolaA & The aggregated traded volume of A/P/M accounts in the VSTOXX front month future. \\
TradVolaP &  \\
TradVolM &  \\
TradVolTot & The aggregated traded volume of accounts in the VSTOXX front month future. \\
AvgFutSpd & The time-weighted average absolute spread of the VSTOXX front month future. \\
AvgFutBidSz & The time-weighted average volume of resting orders on the best bid/ask level of the orderbook. \\
AvgFutAskSz & \\
OptionPosA & The aggregated long/short position of A/P/M accounts in the EURO STOXX 50 options.  \\
OptionPosP & \\
OptionPosM & \\
\end{tabular}
\end{ruledtabular}
\caption{Potential features for a model that explains the difference between market and theoretical value of the VSTOXX front month future contract.}
\label{tab:features}
\end{table*}
\endgroup

\section{Analysis of price deviations}
\subsection{Potential model features}
In the following sections we attempt to explain the deviations between our Heston model and the market-observed VSTOXX future price by means of statistics and machine learning to reveal the driving factors behind them. Constructing a machine learning model first requires the collection of data that potentially could play a role in causing the observed deviations. Naturally, one would expect that the price difference is mainly driven by supply of and demand for the future itself. We have selected a broad list of market features (see table~\ref{tab:features}) related to the VSTOXX future that could potentially play a role, including actual future price, level of the VSTOXX index, number of days to expiry, trader positions, changes in positions, traded volume, average spread and visible resting order size (volume) on the best bid/ask. Positions and traded volumes are included on a per account basis.

\begin{figure*}[t]
\includegraphics[ width=\linewidth ]{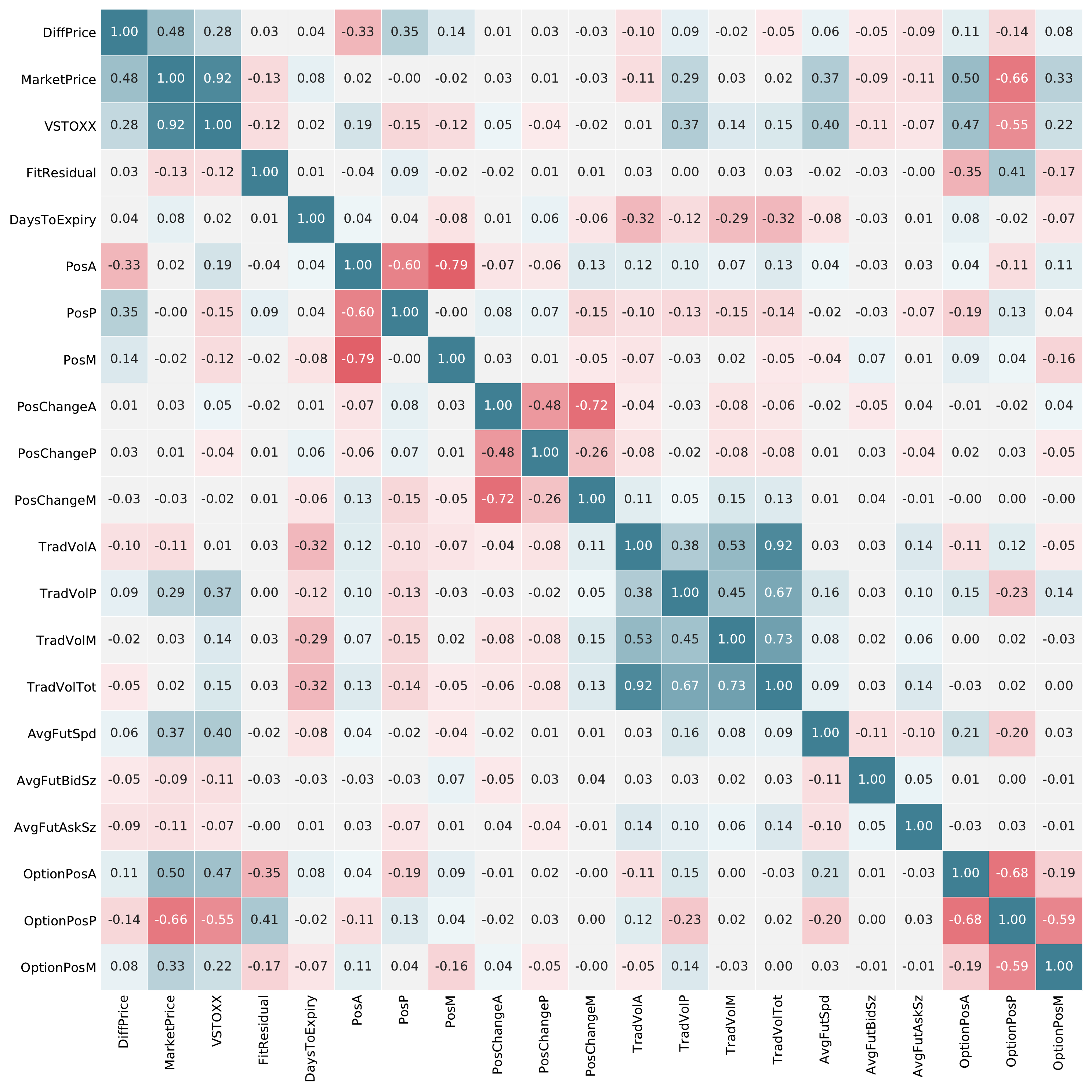}
\caption{Correlation matrix for potential model features that could explain the difference between market and theoretical value of the VSTOXX front month future. The description for each feature is listed in table~\ref{tab:features}. Blue background color indicates a positive correlation, while red background color indicates a negative correlation.}
\label{fig:correlation}
\end{figure*}

At this point we also note that, while the criteria for the A account are clear in the sense that those positions belong to end clients such as investment firms and retailers, the distinction between firms behind P and M accounts is somewhat blurry. The M account is often used by those exchange members which participate in dedicated liquidity provider programs, but the P account may as well be used to trade liquidity providing strategies, even if the respective member does not participate in the exchange's dedicated programs. Nevertheless, the accounts implicitly differentiate between the exchange members active in these products, which may well run different kinds of strategies, since one firm often uses only one type of account for its trading.

We include the aggregated positions in EURO STOXX 50 options, since we expect them to be related to the overall level of the VSTOXX index. We also include the residual from the Heston model fit according to eq.~\ref{eq:mseopt} as a feature, since we need to exclude that the deviation between market and model prices is just an artifact of a failed calibration process, which would be indicated by a strong correlation between price deviation and fit residual. Meanwhile, those features are public market data, since even aggregated position changes are available at least on a daily basis via exchange data feeds. The target variable, namely the price difference, is also included into the list of features for now. Naturally, we will exclude it from the list of features once we get to training the actual machine learning models. All features are available as exactly one data point per trading day. The market prices that enter the model are all settlement prices recorded at 5.30 p.m.~Frankfurt local time, i.e.~CET or CEST. The investigated time frame extends from May 2016 until August 2019.

In a first attempt to understand the relationships between those features, we calculate the correlation coefficients between them. These are displayed in fig.~\ref{fig:correlation} in the form of a matrix with colored entries. Most importantly, the market price of the future, the VSTOXX index and the positions in the A and P accounts are the only features that are significantly correlated with the price difference. A connection between high VSTOXX index/future price and potential price dislocations is sensible, since a high level of the VSTOXX index indicates market stress, which makes price deviations between related instruments more probable.

Furthermore, we observe blocks of correlated features. The per account position, position change and traded volume features are all trivially correlated with the corresponding features for other accounts. The positions and position changes per definition add up to zero, while adding up the traded volume results in the total traded volume. 

More interestingly, both the VSTOXX index and the market price of the VSTOXX future are quite strongly correlated with the trader positions in EURO STOXX 50 options. Such a connection is not apparent for the VSTOXX price and trader positions in the VSTOXX future itself. This actually justifies using our Heston model as a baseline explanation for the VSTOXX future price, since the model is calibrated against the EURO STOXX 50 options, which appear to be the main driver behind the absolute price level of the VSTOXX future. Economically, it is clear that such a relation shall exist, since demand and supply of the EURO STOXX 50 options drive their implied volatility levels, from which the VSTOXX index is derived.

The lack of correlation between the market price of the VSTOXX future and the trader positions in this product suggest that the VSTOXX future should in general be viewed as a dependent product. Therefore, the account positions in the future itself only come into play when explaining the difference between market and theoretical value of the future, since this information is not yet included in our Heston model. The positive correlation between price difference and P account position (in the VSTOXX future) and the negative correlation between price difference and A account position (in the VSTOXX future) means that the market price is usually higher than the theoretical value whenever the P account has a long position and the A account has a short position in the VSTOXX future. This is contrary to the behaviour of the EURO STOXX 50 options market, where client demand seems to drive prices up, as one would expect from basic economic considerations. In the VSTOXX future market it rather seems that occasionally large long positions on P accounts or large short positions on A accounts are driving prices above the theoretical value (compare fig.~\ref{fig:priceposition}). This especially happens around the Brexit vote in 2016 or the French presidential elections in April 2017.

Furthermore, the fit residual is not significantly correlated with any feature except for the option positions, where it seems that the options smile is easier to fit if positions on P accounts are rather in the short and positions on A accounts are in the long region. Most importantly, there is no correlation between price difference and fit residual, which would have pointed to an incorrect model construction in the first place.

For the machine learning models we intend to train in the following sections we consolidate some of the features we included so far. Since the VSTOXX index and the price of the VSTOXX future seem to contain largely the same information, we drop the VSTOXX index and keep the market price of the future. We also drop the EURO STOXX 50 option positions, since the information contained therein is already sufficiently respresented by the VSTOXX futures price and our theoretical model. Finally, we also drop the fit residual, since we concluded that it is not useful in explaining the target variable.

\subsection{Data preparation and software for machine learning}
We now go beyond statistical analysis and enter the territory of machine learning. From now on, we exclude the price difference from the set of features and instead define it as the target variable. We randomly split the data set into a testing set equivalent to 30\% of the data, which is set aside and used purely for verification purposes. No model ever visits those data during the training phase. The other 70\% of the data are used for training the models.

The analysis code is written in the Python programming language and employs the \texttt{scikit-learn} package~\cite{sklearn} for model training and evaluation.

\subsection{Regularized linear model}
We start the investigation with a simple linear model, which is regularized by applying an $L_1$-penalty proportional to a regularization coefficient $\alpha \geq 0$, i.e.~the Lasso.~\cite{lasso} Using this method we extract the coefficients of a linear model relative to the degree of regularization. The regularization reduces overfitting and has two effects on parameters: it decreases the parameter magnitude and makes less important parameters disappear earlier than more important ones.

Before applying the Lasso or the ordinary least squares procedure, we apply standard scaling to all features, i.e.~we subtract the mean value of the feature and then divide by its standard deviation. 

Let us now call the coefficient set obtained from an un-regularized least squares fit $\{\beta_{0,i}\}$, where $i$ corresponds to the i-th feature. The effect of coefficient magnitude shrinkage is measured by the so-called shrinkage factor $s(\alpha)$ defined as:
\beq
s (\alpha) = \frac{\sum\limits_i \beta_{\alpha, i}}{\sum\limits_i \beta_{0, i}}
\eeq
Here, $\beta_{\alpha, i}$ is the $i$-th coefficient determined with a regularization parameter of $\alpha$. Therefore, a shrinkage factor of $s=1$ corresponds to the coefficients of the least-squares estimate, while a shrinkage factor of $s=0$ corresponds to the model that is so strongly regularized that all coefficients have been set to zero.

\begin{figure}[t]
\includegraphics[ width=\linewidth ]{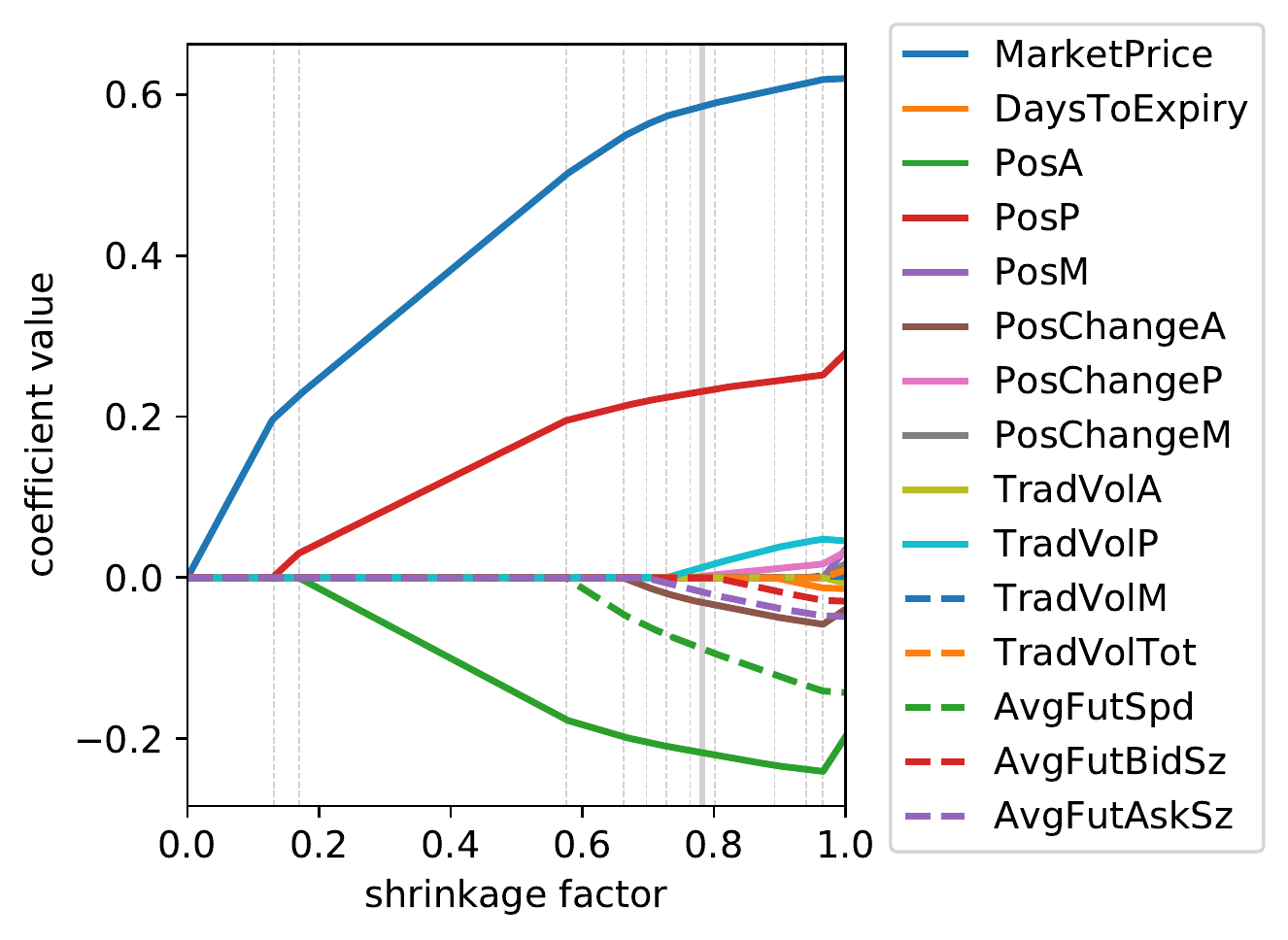}
\caption{Coefficients of the $L_1$-regularized (Lasso) linear model versus shrinkage factor. A shrinkage factor of one corresponds to the non-regularized ordinary least squares estimate. A shrinkage factor of zero corresponds to the regularization coefficient for which all model parameters vanish. Shrinkage factors at which a parameter disappears are marked with a broken vertical grey line. The continuous vertical grey line indicates the optimal shrinkage factor $s=0.782$ chosen by cross-validation. }
\label{fig:lassopath}
\end{figure}

In fig.~\ref{fig:lassopath} we show the so-called Lasso path, i.e.~the coefficients as a function of shrinkage factor. This kind of plot is quite informative with respect to the relative importance of features, since the most influential features disappear last. We observe that the three most important features for describing the price difference are the observed price in the market and positions in the P and A accounts, with about equal magnitude and opposite sign, as expected from the previous correlation analysis. Surprisingly the average bid/ask spread of the future (AvgFutSpd) appears here as the fourth most important feature, which is hard to explain based on a direct correlation with the target variable. However, the correlation matrix (see fig.~\ref{fig:correlation}) shows that the average spread of the future is to some degree collinear with the observed price in the market (MarketPrice), which is the most powerful explanatory variable in our data set. Therefore, the linear model incorrectly assigns a large weight to the average spread of the future. Most of the other features disappear quite early in the regularization process, which points to their relatively low explanatory power.

Finally, we attempt to find the optimal value of $\alpha$ by means of cross-validation. We split the training data set randomly in five different folds into 80\% training and 20\% verification data. For each of those splits we calculate the value of explained variance for a given value of the regularization parameter $\alpha$. We then search for the $\alpha$, which gives the best average explained variance for all of the chosen splits. This procedure yields the model, which represents the optimal compromise between minimal overfit and maximum explanatory power on the given data set. Please note that the testing set we had previously set aside was not used in the cross-validation process. We find that the optimal model is achieved at a shrinkage factor of $s \approx 0.782$. This model still includes several of the less important features. 

The explanatory power of this model can be tested by calculating the explained variance on the testing data set we had set aside initially. The explained variance score on the testing set is $31.8\%$, which is not a particularly high value. Even the explained variance of the training data set is quite low with only $41.3\%$. This either points to the fact that non-linearities are present in the data set or, just as likely, that important features are missing from our list. However, the linear model serves as a benchmark for more complex models.

\subsection{Random forest model}
To adress the potential issue of non-linear relations between features, we also train a random forest~\cite{randomforest} on the training set. Random forests are particularly well suited for working with noisy data sets, since adding additional trees to a random forest does not increase the tendency to overfit, while stabilizing the model with respect to the most important features. We use a random forest built from 250 trees. 

\begin{figure}[t]
\includegraphics[ width=\linewidth ]{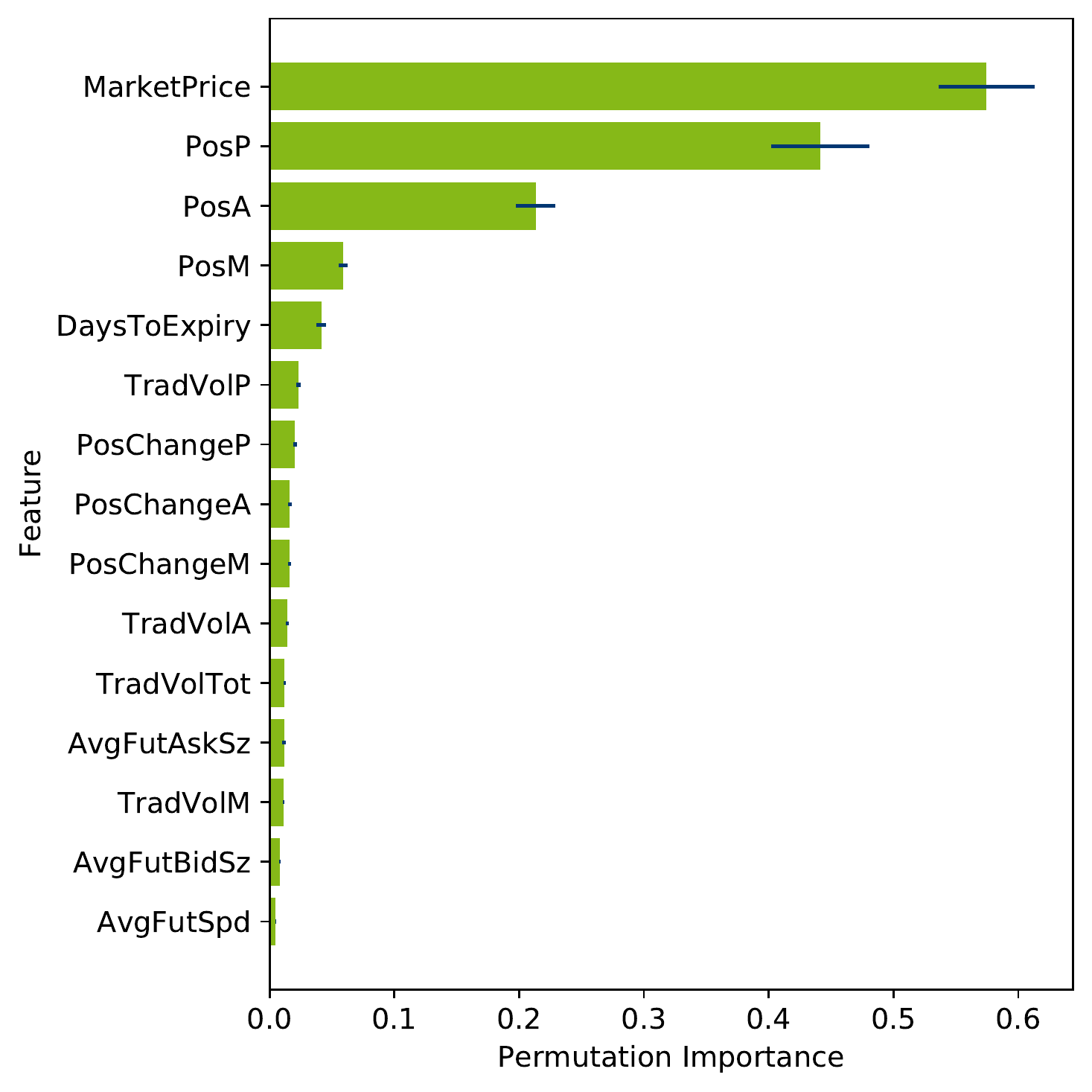}
\caption{Permutation importance of features within the random forest model. The thin blue bars indicate the standard deviation of permutation importance for the given feature.  MarketPrice, PosP and PosA are associated with the highest values of permutation importance, i.e.~they contribute the greatest amount of explanatory power to the model.}
\label{fig:featureimp}
\end{figure}

The importance of features in the trained model can be measured by the concept of permutation importance~\cite{permutationimportance}. Permutation importance quantifies the mean decrease in accuracy following the removal of a single feature from the model. For estimating the permutation importance, we train 30 different random forests on the same data by varying the initial random state and extract the importance of each feature using the data obtained from the 30 trained random forests. We then calculate the mean value of importance and its standard deviation for each feature. The results are shown in fig.~\ref{fig:featureimp}.

The permutation importance measure clearly singles out MarketPrice, PosP and PosA as the most important features in the random forest. For all features the standard deviation of permutation importance is small compared to its average value. Therefore, the estimates of permutation importance can be considered reliable. The next most important features are PosM and DaysToExpiry. From a business perspective it makes sense that positions on market making accounts should also have some influence. The number of days to expiry may also play a role, since price deviations should get smaller as the product approaches its expiry date. The rest of the features play only a minor role and can be considered practically irrelevant. In particular, the feature AvgFutSpd is rated as least relevant by the random forest, while it was quite prominent in the linear model (see fig.~\ref{fig:lassopath}).

\begin{figure*}[t]
\includegraphics[ width=\linewidth ]{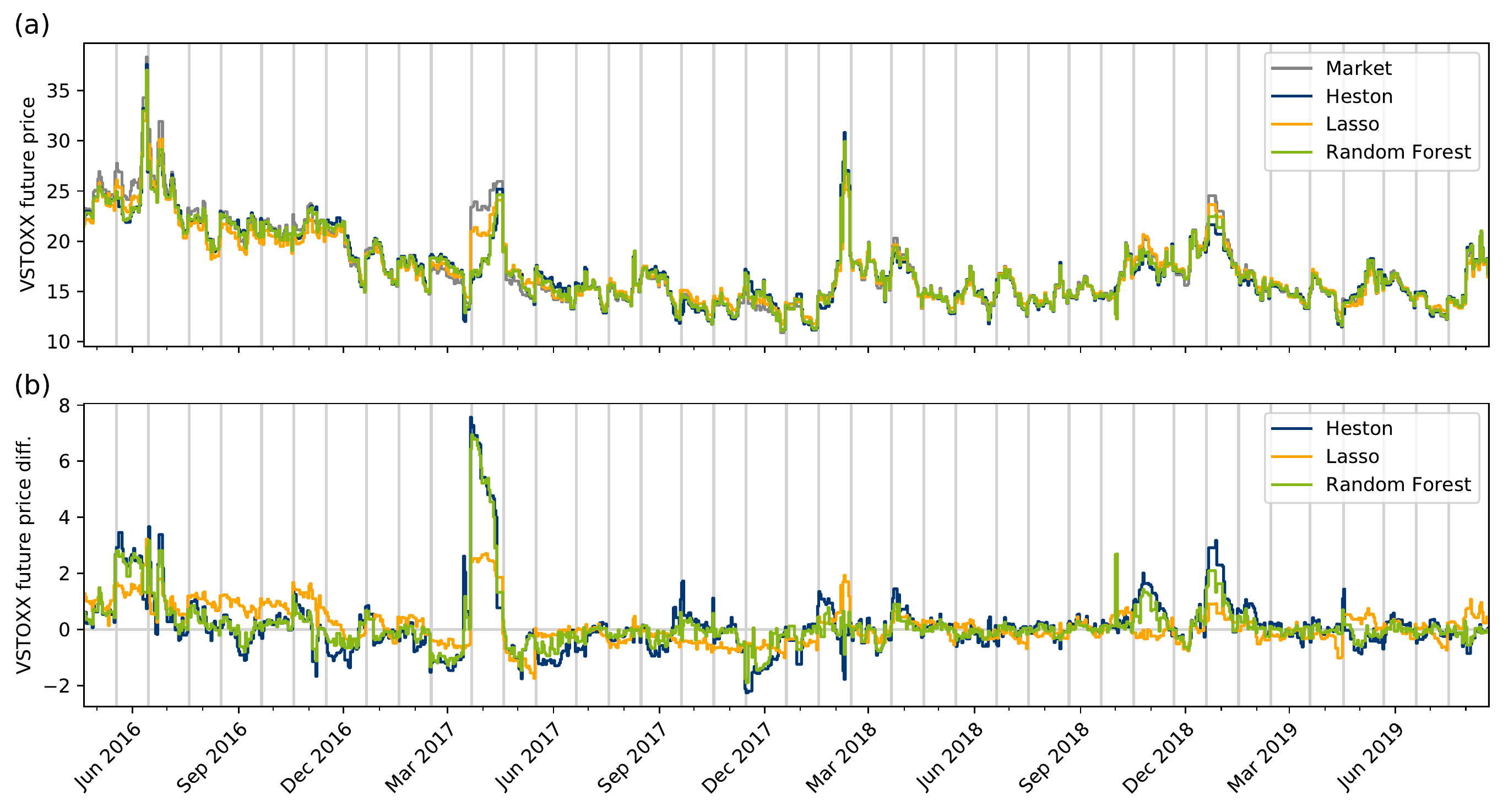}
\caption{(a) Market and model front month VSTOXX future prices. (b) Difference between market and model prices. The regularized linear model (Lasso) is only able to explain a fraction of the observed data points, while the random forest reproduces most of the observed price deviations.}
\label{fig:modelresults}
\end{figure*}

The explanatory power of the random forest model is tested by applying it to the testing data we had set aside initially. The explained variance on the testing set is $70.2\%$, while the explained variance on the training set is $96.2\%$. These scores are significantly higher than for the linear model. The higher explained variance score on the training set is not surprising, since the random forest has a much larger number of parameters and can practically fit every detail of the training data set. The explained variance score of $70.2\%$ is a large improvement compared to the result for the linear model. Although the random forest does not capture all details of the test data set, its main features are explained reasonably well. In fig.~\ref{fig:modelresults} we compare the theoretical prices obtained by both the Heston and our machine-learning augmented models. The figure clearly confirms that the random forest delivers a good explanation for the observed deviations between market price and Heston model price.

\subsection{Interpretation}
We have found that the VSTOXX future price is usually consistent with the EURO STOXX 50 option market. Since the positions in the VSTOXX future itself are not correlated with the VSTOXX index, we can safely conclude that the EURO STOXX 50 options are driving the VSTOXX index and, consequently, the price of the VSTOXX future. Only on rare occasions, usually linked to major political events, do the positions in the VSTOXX future actually drive its price. Interestingly, large short positions of A accounts are driving the market price above the theoretical value. The fact that clients selling drives the price of a product upwards at first glance seems to contradict the usual law of supply and demand. 

For the EURO STOXX 50 options we had observed that client demand drives implied volatility upwards, as one would expect. Since a long VSTOXX future is commonly used to hedge short volatility positions in EURO STOXX 50 options, it seems logical that the role of A and P accounts could be reversed when it comes to supply and demand in the VSTOXX future compared to the EURO STOXX 50 options. This insight explains why large short positions of the A account in VSTOXX futures are correlated with a price above the theoretical value. 

\section{Conclusions}
We have constructed a no-arbitrage pricing model for the VSTOXX future based on the Heston stochastic volatility model, which is calibrated against EURO STOXX 50 options and the VSTOXX index. We used this model to calculate theoretical prices for the VSTOXX future. While we observed that theoretical and market prices agree reasonably well for most of the observed time period from May 2016 to August 2019, large deviations are visible over a few extended time spans close to major politcal events. From this we conclude that the price of the VSTOXX future is usually tied to the EURO STOXX 50 options market, which is plausible, since the underlying VSTOXX index is calculated based on EURO STOXX 50 option prices.

For the deviations between observed market price and Heston model price we attempted to find an explanation based on trader account positions and other features of the market. We collected a time series of these features and performed a correlation analysis. Using the insights gained from the correlations, we trained a regularized linear model and a random forest based on the collected data. We find that both models select the market price of the future itself and the accumulated positions of different types of traders as the features with most explanatory power with respect to the observed price deviations. The predictions of the non-linear random forest perform significantly better than the linear model, which points to non-linear relationships between the important variables in the problem.

Although we have focused on explaining price differences, our machine learning model could also be used in combination with the Heston model to reproduce the market price to a higher degree of accuracy. We expect that models along these lines, although probably more involved, are used by market participants.

We also observed an inversion of account roles with respect to the law of supply and demand between the EURO STOXX 50 options and the VSTOXX future. We attribute this inversal to the fact that these products are used to hedge each other, with the EURO STOXX 50 options dominating this relationship.

In summary, we find that aggregated trader position information explains most of the observed deviations between market and theoretical price of the VSTOXX future between May 2016 and August 2019. Since similar aggregated information have recently become available as market data streams, we expect that market participants will use models comparable to the one presented here to further improve the pricing efficiency in the VSTOXX future and other markets.

\begin{acknowledgments}
The author thanks Sascha Semroch for valuable discussions on the VSTOXX future as well as Michail Koslov for sharing his insights on various machine learning techniques. He also thanks Zubin Ramdarshan for proofreading the manuscript.
\end{acknowledgments}



\begin{thebibliography}{9}

\bibitem{carr2009}
P. Carr and R. Lee,
\textit{Volatility Derivatives},
Annu. Rev. Financ. Econ. \textbf{1}, 319 (2009), 
\doi{10.1146/annurev.financial.050808.114304}.

\bibitem{gruenbichler1996}
A. Grünbichler and F. A. Longstaff,
\textit{Valuing Futures and Options on Volatility},
J. Bank. Finance \textbf{20}, 985 (1996),
\doi{10.1016/0378-4266(95)00034-8}.

\bibitem{zhang2006}
J. Zhang and Y. Zhu,
\textit{VIX Futures},
J. Futures Mark. \textbf{26}, 521 (2006),
\doi{10.1002/fut.20209}.

\bibitem{lin2007}
Y.-N. Lin,
\textit{Pricing VIX futures: Evidence from integrated physical and risk‐neutral probability measures},
J. Futures Mark. \textbf{27}, 1175 (2007),
\doi{10.1002/fut.20291}.

\bibitem{zhu2007}
Y. Zhu and J. Zhang,
\textit{Variance term structure and VIX futures pricing},
Int. J. Theor. Appl. Finance \textbf{10}, 11 (2007),
\doi{10.1142/S0219024907004123}.

\bibitem{sepp2008}
A. Sepp,
\textit{VIX Option Pricing in a Jump-Diffusion Model},
Risk Magazine \textbf{21}, 84 (2008).

\bibitem{lin2009}
Y.-N. Lin and Ch.-H. Chang,
\textit{VIX option pricing},
J. Futures Mark. \textbf{29}, 523 (2009),
\doi{10.1002/fut.20387}.

\bibitem{lu2009}
Z. Lu and Y. Zhu,
\textit{Volatility components: The term structure dynamics of VIX futures},
J. Futures Mark. \textbf{30}, 230 (2009),
\doi{10.1002/fut.20415}.

\bibitem{psychoyios2010}
D. Psychoyios, G. Dotsis and R. N. Markellos,
\textit{A jump diffusion model for VIX volatility options and futures},
Rev. Quant. Finan. Acc. \textbf{35}, 245 (2010),
\doi{10.1007/s11156-009-0153-8}.

\bibitem{duan2010}
J.-Ch. Duan and Ch.-Y. Yeh,
\textit{Jump and volatility risk premiums implied by VIX},
J. Econ. Dyn. Control \textbf{34}, 2232 (2010),
\doi{10.1016/j.jedc.2010.05.006}.

\bibitem{lin2010}
Y.-N. Lin and Ch.-H. Chang,
\textit{Consistent modeling of S\&P 500 and VIX derivatives},
J. Econ. Dyn. Control \textbf{34}, 2302 (2010),
\doi{10.1016/j.jedc.2010.02.003}.

\bibitem{zhang2010}
J. E. Zhang, J. Shu, and M. Brenner,
\textit{The new market for volatility trading},
J. Futures Mark. \textbf{30}, 809 (2010),
\doi{10.1002/fut.20448}.

\bibitem{wang2011}
Z. Wang and R. T. Daigler,
\textit{The performance of VIX option pricing models: Empirical evidence beyond simulation},
J. Futures Mark. \textbf{31}, 251 (2011),
\doi{10.1002/fut.20466}.

\bibitem{zhu2011}
S. Zhu and G. Lian,
\textit{A closed-form exact solution for pricing variance swaps with stochastic volatility},
Mathematical Finance \textbf{21}, 233 (2011),
\doi{10.1111/j.1467-9965.2010.00436.x}.

\bibitem{zhulian2012}
S.-P. Zhu and G.-H. Lian,
\textit{An analytical formula for VIX futures and its applications},
J. Futures Mark. \textbf{32}, 166 (2012),
\doi{10.1002/fut.20512}.

\bibitem{luo2012}
X. Luo and J. E. Zhang,
\textit{The Term Structure of VIX},
J. Futures Mark. \textbf{32}, 1092 (2012),
\doi{10.1002/fut.21572}.

\bibitem{hancock2012}
G. D'Anne Hancock,
\textit{VIX and VIX Futures Pricing Algorithms: Cultivating Understanding},
Modern Economy \textbf{3}, 284 (2012),
\doi{10.4236/me.2012.33038}.

\bibitem{lian2013}
G.-H. Lian and S.-P. Zhu,
\textit{Pricing VIX options with stochastic volatility and random jumps},
Decis. Econ. Finance \textbf{36}, 71 (2013),
\doi{10.1007/s10203-011-0124-0}.

\bibitem{cont2013}
R. Cont and Th. Kokholm,
\textit{A Consistent Pricing Model for Index Options and Volatility Derivatives},
Mathematical Finance \textbf{23}, 248 (2013),
\doi{10.1111/j.1467-9965.2011.00492.x}.

\bibitem{baldeaux2014}
\textit{Consistent Modelling of VIX and Equity Derivatives Using a 3/2 plus Jumps Model},
Appl. Math. Finance \textbf{21}, 299 (2014),
\doi{10.1080/1350486X.2013.868631}.

\bibitem{frijns2016}
B. Frijns, A. Tourani-Rad, and R. I. Webb,
\textit{On the Intraday Relation Between the VIX and its Futures},
J. Futures Mark. \textbf{36}, 870 (2016),
\doi{10.1002/fut.21762}.

\bibitem{li2017}
J. Li, L. Li, and G. Zhang,
\textit{Pure jump models for pricing and hedging VIX derivatives},
J. Econ. Dyn. Control \textbf{74}, 28 (2017),
\doi{10.1016/j.jedc.2016.11.001}.

\bibitem{pacati2018}
C. Pacati, G. Pompa, and R. Ren\`o,
\textit{Smiling twice: The Heston++ model},
J. Bank. Finance \textbf{96}, 175 (2018),
\doi{10.1016/j.jbankfin.2018.08.010}.

\bibitem{lo2019}
Ch.-L. Lo, P.-T. Shih, Y.-H. Wang, and M.-T. Yu,
\textit{VIX derivatives: Valuation models and empirical evidence},
Pac.-Basin Financ. J. \textbf{53}, 1 (2019),
\doi{10.1016/j.pacfin.2018.09.004}.

\bibitem{heston1993}
S. L. Heston,
\textit{A Closed-Form Solution for Options with Stochastic Volatility with Applications to Bond and Currency Options},
Rev. Financ. Stud. \textbf{6}, 327 (1993),
\doi{10.1093/rfs/6.2.327}.

\bibitem{blackscholes73}
F. Black and M. Scholes,
\textit{The Pricing of Options and Corporate Liabilities},
J. Political Econ. \textbf{81}, 637 (1973),
\doi{10.1086/260062}.

\bibitem{jaeckelrational}
P. Jäckel,
\textit{Let's be rational},
Wilmott Magazine \textbf{1}, 40 (2015),
\doi{10.1002/wilm.10395}.

\bibitem{lefloch2018}
F. Le Floc'h,
\textit{An adaptive Filon quadrature for stochastic volatility models},
J. Comput. Finance \textbf{22}, 65 (2018),
\doi{10.21314/JCF.2018.356}.

\bibitem{guterding2018}
D. Guterding and W. Boenkost,
\textit{The Heston stochastic volatility model with piecewise constant parameters - efficient calibration and pricing of window barrier options},
J. Comput. Appl. Math. \textbf{343}, 353 (2018),
\doi{10.1016/j.cam.2018.04.054}.

\bibitem{lorig2017}
M. Lorig, S. Pagliarani, and A. Pascucci,
\textit{Explicit Implied Volatilities for Multifactor Local‐Stochastic Volatility Models},
Mathematical Finance \textbf{27}, 926 (2017),
\doi{10.1111/mafi.12105}.

\bibitem{diffevo}
R. Storn and K. Price,
\textit{Differential Evolution - a Simple and Efficient Heuristic for Global Optimization over Continuous Spaces},
J. Global Optim. \textbf{11}, 341 (1997),
\doi{10.1023/A:1008202821328}.

\bibitem{lbfgs}
R. H. Byrd, P. Lu, J. Nocedal, and C. Zhu,
\textit{A Limited Memory Algorithm for Bound Constrained Optimization},
SIAM J. Sci. Comput. \textbf{16}, 1190 (1995),
\doi{10.1137/0916069}.

\bibitem{sklearn}
F. Pedregosa, G. Varoquaux, A. Gramfort, V. Michel, B. Thirion, \textit{et al.},
\textit{Scikit-learn: Machine Learning in Python},
J. Mach. Learn. Res. \textbf{12}, 2825 (2011),
\doi{10.5555/1953048.2078195}.

\bibitem{lasso}
R. Tibshirani,
\textit{Regression shrinkage and selection via the lasso},
J. Royal. Statist. Soc. B \textbf{58}, 267 (1999),
\doi{10.1111/j.2517-6161.1996.tb02080.x}.

\bibitem{randomforest}
L. Breiman,
\textit{Random Forests},
Machine Learning \textbf{45}, 5 (2001),
\doi{10.1023/A:1010933404324}.

\bibitem{permutationimportance}
A. Altmann, L. Tolosi, O. Sander, and T. Lengauer,
\textit{Permutation importance: a corrected feature importance measure},
Bioinformatics \textbf{26}, 1340 (2010),
\doi{10.1093/bioinformatics/btq134}.

\end{thebibliography}
\end{document}